\documentclass[final,3p,times]{elsarticle}

\usepackage{amssymb}
\usepackage{float}
\usepackage{amsmath}
\usepackage{caption}
\usepackage[inkscapelatex=false]{svg}
\usepackage{pdfpages}
\usepackage{siunitx}
\usepackage{natbib}
\usepackage{tabularx}
\usepackage{booktabs}
\usepackage{graphicx}
\usepackage{booktabs}
\usepackage{multirow}
\usepackage{chngcntr}
\usepackage{adjustbox}
\usepackage{ulem}
\journal{arXiv}
\begin{document}
\begin{frontmatter}
\title{TopoGEN: topology-driven microstructure generation 
    \\ for in silico modeling of fiber network mechanics}

\author[label1]{Sara Cardona}
\author[label1]{Mathias Peirlinck\corref{cor1}}
\ead{mplab-me@tudelft.nl}
\author[label1]{Behrooz Fereidoonnezhad\corref{cor1}}
\ead{b.fereidoonnezhad@tudelft.nl}
\cortext[cor1]{Senior authors contributed equally. Correspondence:}
\affiliation[label1]{organization={Department of BioMechanical Engineering, Faculty of Mechanical Engineering, Delft University of Technology},
            country={the Netherlands}}
            
\begin{abstract}
The fields of mechanobiology and biomechanics are expanding our understanding of the complex behavior of soft biological tissues across multiple scales. 
Given the intricate connection between tissue microstructure and its macroscale mechanical behavior, unraveling this mechanistic relationship remains an ongoing challenge.
Reconstituted fiber networks serve as valuable in vitro models to simplify the intricacy of in vivo systems for targeted investigations.
Concurrently, advances in imaging enable microstructure visualization and, through generative pipelines, modeling as discrete element networks. 
These mesoscale (\textmu m) models provide insights into macroscale (mm) tissue behavior.
However, there is still no clear way to systematically incorporate detailed experimentally observed microstructural changes into in silico models of biological networks.
In this work, we develop a novel framework to generate topologically-driven discrete fiber networks using high-resolution images that account for how environmental changes during polymerization influence the resulting structure.
Leveraging these networks, we generate models of interconnected load-bearing fiber components that exhibit softening under compression and are bending-resistant.  
The generative topology framework enables control over network-level features, such as fiber volume fraction and cross-link density, along with fiber-level properties, like length distribution, to simulate changes driven by different polymerization conditions.
We validate the robustness of our simulations against experimental data in a collagen-specific study case where we examine nonlinear elastic responses of collagen networks across varying conditions.
TopoGEN provides a versatile tool for tissue biomechanics and engineering, helping to bridge microstructural insights and bulk mechanical behavior by linking image-derived microstructural topological organization to soft tissue mechanics.
\end{abstract}


\end{frontmatter}

\section{Introduction}
\label{Intro}
Microstructure drives function. 
In soft biological materials, the microstructure comprises a complex fibrous network known as the extracellular matrix. 
This intricate three-dimensional meshwork of fibers provides mechanical stability, elasticity, and strength and plays a crucial role in governing cellular processes \citep{Dean2023}.
Cells interact with the extracellular matrix through mechanotransduction \citep{Jansen2015}, a process by which they sense environmental changes and convert mechanical impulses into biological responses \citep{Fereidoonnezhad2017, Peirlinck2019}. 
The discipline of mechanobiology revolves around this dynamic feedback loop \citep{Loerakker2022}, which is regulated by highly localized micromechanical factors. 
A key challenge, shared across many engineering applications \citep{Geers2010}, is understanding the relationship between microstructural features (e.g., individual constituents and network properties) and macroscopic functions (e.g., mechanical behavior) of these networks.
In the specific case of soft biological tissues, the primary challenge lies in the geometrical complexity and weakly connected microstructure that disrupts deformation affinity and gives rise to localized deformations \citep{Mirzaali2020} highlighted in experimental studies \citep{Cavinato2020}.
Therefore, in biological tissues, the relative spatial position of two neighboring particles after deformation cannot be fully described solely by their relative material position before deformation, and non-affinity must be considered.
Addressing this challenge requires an in-depth investigation of details associated with the individual constituents of the networks, such as fiber kinematics \citep{Mahutga2023}.
Early computational models of soft biological tissues were predominantly continuum-based, drawing inspiration from histological data \citep{Fung1967, Holzapfel2000, Gasser2005}. 
These models rely on intuitive understanding, data availability, and a priori made physical assumptions. 
Thus, they serve as descriptive tools of material mechanics but fall short in predicting complex loading scenarios or new material behavior \citep{Stylianopoulos2007, dissertation, Picu2021, Sun2021, Mahutga2023}. 
Most notably, they fail to offer mechanistic insights into material behavior based on the structural information of their building blocks as existing continuum models cannot capture the non-affine fiber kinematics, resulting in predictions that are too stiff compared to experiments \citep{Picu2021}.
Advancements in imaging techniques have led to more detailed representations of the extracellular space through network models.
Such network models explicitly depict interconnected fibers as discrete networks \citep{Lindstroem2010, Picu2011, Lindstroem2013, Nan2018, Ban2019, Merson2020, Dalbosco2021, Eichinger2021, Leng2021, Nikpasand2021, Filla2023, Kakaletsis2023, Mahutga2023, Wahlsten2023}, establishing the connection between fiber mechanics and tissue behavior.
On the experimental side, tissue engineers have developed quantitative techniques to precisely control the microstructure of reconstituted gels by adjusting temperature or pH \citep{Roeder2002, Raub2007, Raub2008, Hwang2011, Wolf2013, Sapudom2015, Morozova2018, McCluskey2020, Oh2020, GonzalezMasis2020, Sun2021, Tan2024} and assess their influence on mechanical response \citep{Jansen2018}.
These analyses show that even small environmental changes can significantly impact microstructure.
The bulk mechanical response results from multiple coexisting factors.
Potential microstructural factors include fiber interconnectivity (i.e., the topology), fiber concentration in the sample volume, and their morphological and mechanical properties.
However, the interrelated and multiscale nature of these factors makes it challenging to disentangle the influence of individual variables on the overall mechanical behavior of the sample.
Computational models that adopt discrete fiber networks offer a highly valuable complementary approach to experiments, enabling a systematic examination of how specific microstructural features contribute to macroscopic properties. 
Most of the current in silico discrete fiber-based approaches are concentration-driven rather than topology-driven, matching fiber volume fractions from experiments. 
However, this approach overlooks the critical role of network topology and other microstructural parameters in shaping mechanical behavior.
Understanding the localized deformations and varied mechanical properties of these reconstituted biological networks is crucial for uncovering how mechanical cues regulate cellular behavior, influence tissue development, and contribute to disease progression \citep{Putten2016, Herum2017, Ayad2019,Buskermolen2020, Eichinger2021a, Espina2021, Kim2021, Atcha2023, Kumar2024, McEvoy2024}.
Moreover, investigating how the microstructure, in terms of fiber and network properties, relates to the measured properties enables material behavior prediction and provides a solid foundation for designing engineered tissues with tailored macroscopic mechanics \citep{Burla2019, Burla2020}.

In this study, we propose TopoGEN, a framework that integrates three-dimensional image-informed fiber network generation with non-linear finite element analysis to support the mechanistic investigation of structure-function relationships in soft matter. 
Our framework features a novel generative pipeline designed to create topologically rich and microstructurally diverse discrete fiber networks that replicate architectural variations induced by experimental conditions (for example, temperature or pH).
As a case study, we examine how microstructural changes in simulated collagen gels influence macroscale mechanical behavior. 
We compare these simulations to experimentally studied reconstituted collagen networks, where similar topological variations are induced by polymerization temperature \citep{Jansen2018}.
Unlike previous studies \citep{Stylianopoulos2007, Jansen2018, Eichinger2021, Leng2021, Deogekar2019a, Deogekar2019, Picu2023, Kakaletsis2023}, which varied individual parameters such as average connectivity at fixed or random fiber lengths, or concentration at fixed length and connectivity, TopoGEN enables simultaneous control of all these structural features. 
This comprehensive control supports a deeper understanding of how the complete microstructural organization affects the bulk mechanical response of soft biological materials.
Specifically, to extend the analysis beyond experimentally accessible mesoscale parameters like concentration and connectivity, we leverage TopoGEN’s ability to generate microstructures with targeted variations, altering one feature while keeping others constant, to isolate the individual effects of key microstructural attributes, including fiber morphology and constituent mechanics.
This capability is achieved through explicit control over critical parameters, including fiber connectivity, morphology, and concentration, during the generative process, as demonstrated in prior works \cite{Nan2018, Lindstroem2010, Lindstroem2013, Eichinger2021}. 
However, unlike these earlier approaches, TopoGEN imposes biologically motivated constraints on network architecture. 
It limits the number of connections per node to represent loosely connected structures where fibers interact only via cross-links or branches, excluding configurations with higher-order connectivity that are less representative of biological reality.
\section{Methods}
\label{Methods}

\subsection{Fiber network generation}
\label{GenPipeline}
We model biological networks as structures of connected fibers. 
To generate these networks, we partition a three-dimensional cubic domain into randomly seeded Voronoi polyhedra and represent the fibers and cross-links as the tessellation and vertices, respectively. 
We then optimize this Voronoi-based fiber network using an iterative optimization technique to match key topological features, i.e., average connectivity and fiber length distribution, to those of experimentally measured in biological fiber networks.

\begin{figure}[H]
\centerline{\includegraphics[width=\textwidth]{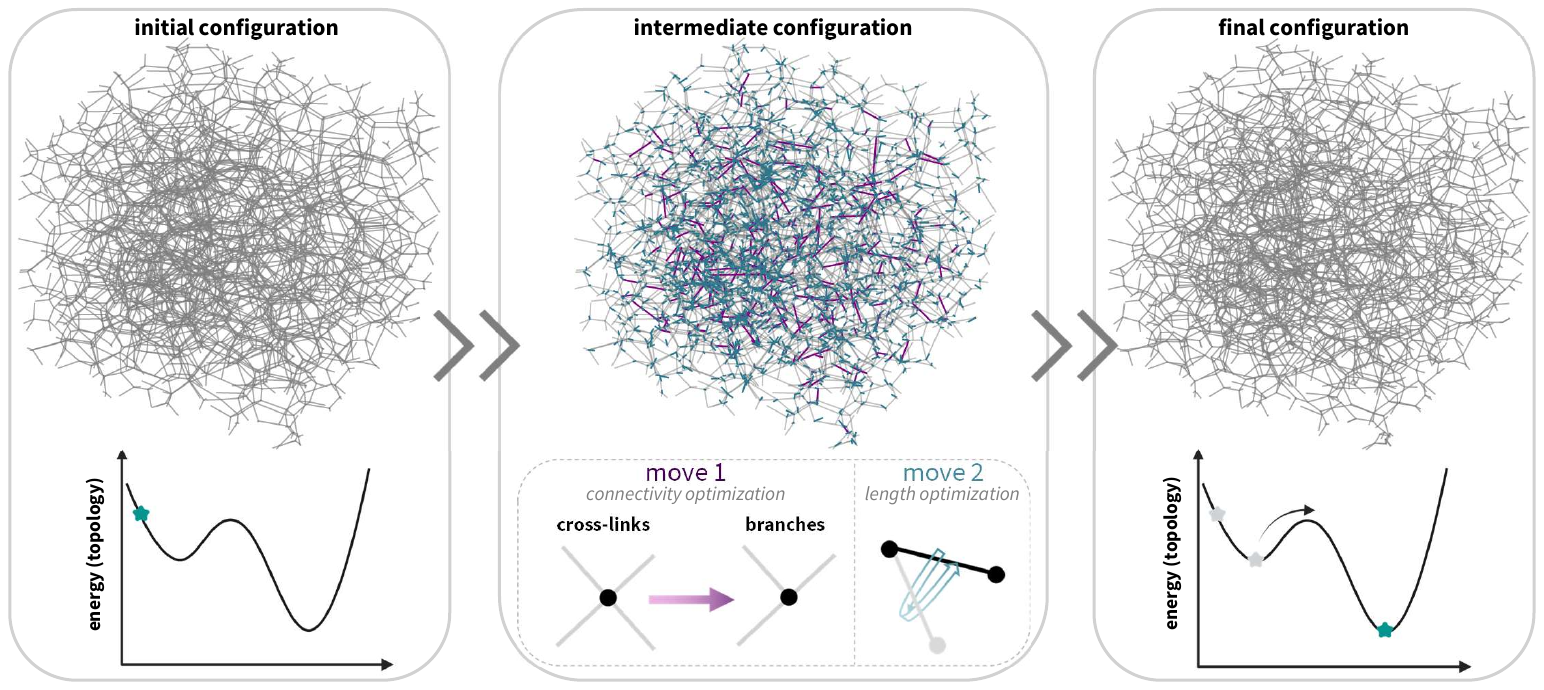}}
\caption{\textbf{Optimization workflow.} 
Starting from an initial random configuration, the algorithm iteratively minimizes the system's energy by removing fibers (move 1) or moving nodes (move 2) until the target distribution of connectivity and fiber length is achieved.}
\label{fig: SimulatedAnnealing} 
\end{figure}

To transform our initial Voronoi-generated random fiber network into neighboring states, we introduce two types of moves illustrated in Figure \ref{fig: SimulatedAnnealing}: dilutive transformations, which randomly remove fibers from the network, and density-preserving transformations, which randomly move nodes within the cubic domain.
We define cross-links as nodes with four-fold connectivity and branches as nodes with three-fold connectivity.
Starting from the four-fold connectivity typical of Voronoi networks, dilution allows branch formation until the target connectivity is reached.
The Poisson length distribution used to generate our initial network results in a number of short fibers, which are inconsistent with the experimentally measured target distribution. 
To address this, we assign higher removal probabilities to shorter edges during the dilutive transformations and iteratively adjust node positions to match the log-normal distribution during the density-preserving transformations.
Once the target connectivity is achieved, we apply a simulated annealing procedure \citep{Yeong1998, Lindstroem2010, Lindstroem2013, Eichinger2021} to optimize the fiber length distribution.
We set the to-be-minimized annealing system energy as the distance of the initial microstructural configuration from the target microstructural configuration achieved by gradually cooling the network and allowing it to transition to new neighboring microstructures. 
These neighboring states are generated by applying small random perturbations to the configuration at each iteration until the current and target distributions match.
To address the challenge of directly comparing continuous length distributions to the log-normal target \citep{Lindstroem2010}, we implement a binning algorithm that partitions fiber lengths into disjoint intervals ($b$).
Therefore, we assign each length observation ($x$) to a specific interval and compute a discretized cost function expressing the difference between the binned distribution and the target. 
To this aim, we implement the Kullback–Leibler (KL) divergence \citep{Kullback1951}, also known as relative entropy ($\mathcal{D}_{KL}$), which is a non-symmetric measure quantifying how one probability distribution $p(x)$ diverges from a second reference probability $q(x)$ distribution:
\begin{equation}
\mathcal{D}_{KL}(\, p(x)\parallel q(x)) =  \sum_{j=1}^{b} p(x_{j}) \log \left( \frac{p(x_{j})}{q(x_{j})} \right)
\label{eq: entropy}
\end{equation}
At each iteration, moves are accepted based on the Metropolis criterion \citep{Metropolis1953}, with a temperature-like parameter \( T \) controlling the likelihood of accepting worse configurations. We set the temperature to decrease exponentially as \( T = 0.95^k \cdot T_0 \), where \( k \) is the iteration step \citep{Eichinger2021, Nan2018}. The initial temperature \( T_0 \) is set to accept ~50\% of moves that increase the energy, as in \citep{Eichinger2021, Nan2018}. Annealing stops when \( T < 1 \times 10^{-4} \cdot T_0 \), when the relative energy drop stays below \( 1 \times 10^{-5} \) over 500 iterations, or when the maximum number of iterations is reached.

Our generative pipeline relies on four user-defined physical variables informed by the microstructure: fiber radius, target length distribution, average connectivity, and the volume of the RVE.
The volume fraction, defined as the ratio of total fiber volume to total RVE volume, is a dependent variable influenced by these four parameters. 
To control the volume fraction, the user can adjust the number of seeds for Voronoi tessellation. 
To automate the selection of the optimal seed count that matches the target volume fraction, we propose a simple method. Since length optimization preserves density and only slightly affects volume fraction, it is reasonable to infer that connectivity optimization provides a reliable indication of how volume fraction responds to changes in the number of seeds.
Starting from a user-provided seed estimate, a binary search algorithm determines the seed count that best approximates the desired volume fraction before launching the generative pipeline.
Supplementary materials include additional analysis regarding the relationship between the fiber length and the number of seeding points (see Figure \ref{fig: seeds_length}).

\subsection{Mechanical equilibrium at the mesoscale}
\label{Equilibriums}
Modeling soft tissues at the level of individual fibers (microscale) or the level of discrete fiber networks (mesoscale) is computationally impractical for large volumes undergoing complex loading and boundary conditions. 
Instead of extending our discrete fiber network model to the entire tissue scale (macroscale), we focus on a representative volume element (RVE) large enough to statistically capture the mechanical properties of the microstructure yet small enough to reduce the computational cost.
Within this mesoscale domain, we define a constitutive behavior for each individual fiber in the network and apply macroscopic deformation gradients (i.e., tissue loading conditions) as boundary conditions.
Solving this boundary value problem with the finite element method yields the macroscale's local (i.e., for every finite element, at each integration point, within every time step, at each Newton iteration) constitutive law \citep{Geers2010, Peirlinck2024}.
In this study, we use the commercial finite element analysis software Abaqus \cite{Abaqus} to retrieve the numerical solution of the RVEs boundary value problem.
The transition from the RVE to the full tissue scale is only admissible if the Hill-Mandel condition \citep{Hill1963} is satisfied, ensuring that the potential energy remains consistent when passing from the meso- to the macroscale.
This condition is ideally met with infinitely large RVEs; however, with finite-sized RVEs, the consistency requires careful handling of the boundary conditions and the RVE size.
This can be achieved through three approaches: (1) enforcing zero microfluctuations throughout the RVE, i.e., coupling local RVE deformation affinely to the macroscale deformation gradient;  (2) setting zero microfluctuations only at the boundary via uniform displacement boundary conditions while allowing internal nodes to deform non-affinely  (i.e., with deviations from a direct one-to-one correspondence to the macroscale deformation gradient); or (3) imposing periodicity of the microfluctuation field at the boundary via periodic boundary conditions (PBCs) and allowing internal nodes to deform non-affinely. 
In this work, we adopt PBCs since they are known to estimate the overall material properties better than the other alternatives mentioned \citep{Miehe2002,Sluis2000,Terada2000,Kanit2006, Peric2010} and lie closer to the effective properties compared to the upper and lower bounds defined by uniform displacement and traction boundary conditions.
Another advantage of adopting PBCs is their ability to capture the effect of non-affinity on the RVE response by allowing internal nodes to move freely within our loosely connected networks.
Given that biological networks are - in essence - not periodic structures, incorporating these PBCs is not a trivial task. 

\begin{figure}[H]
\includegraphics[width=\textwidth]{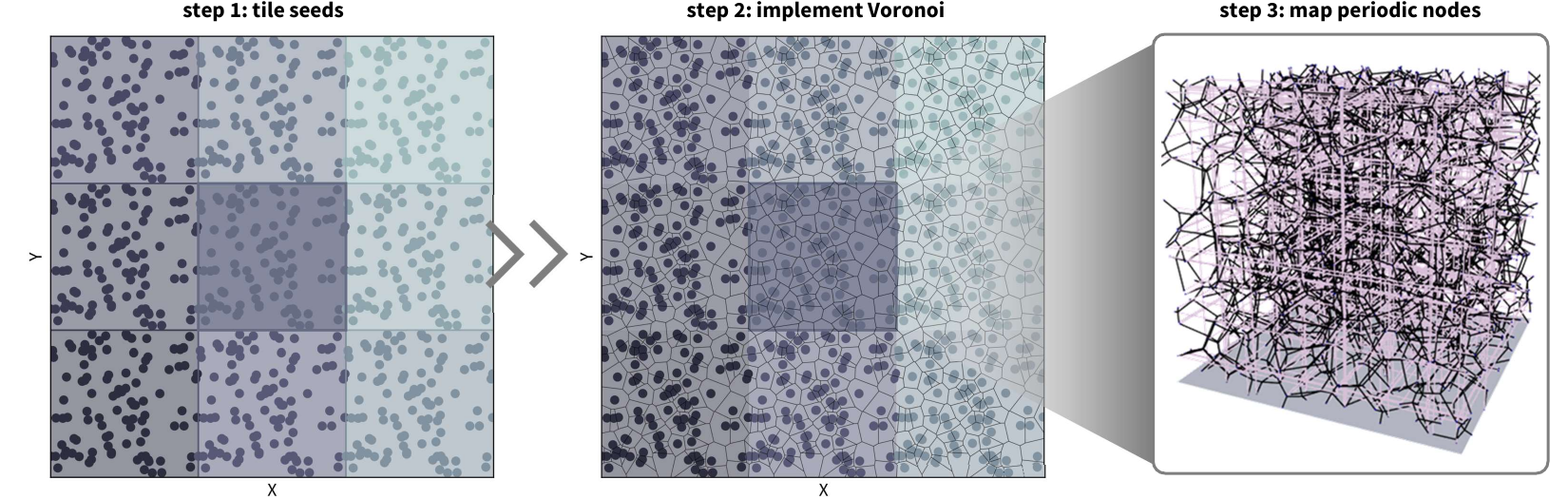}
\includegraphics[width=\textwidth]{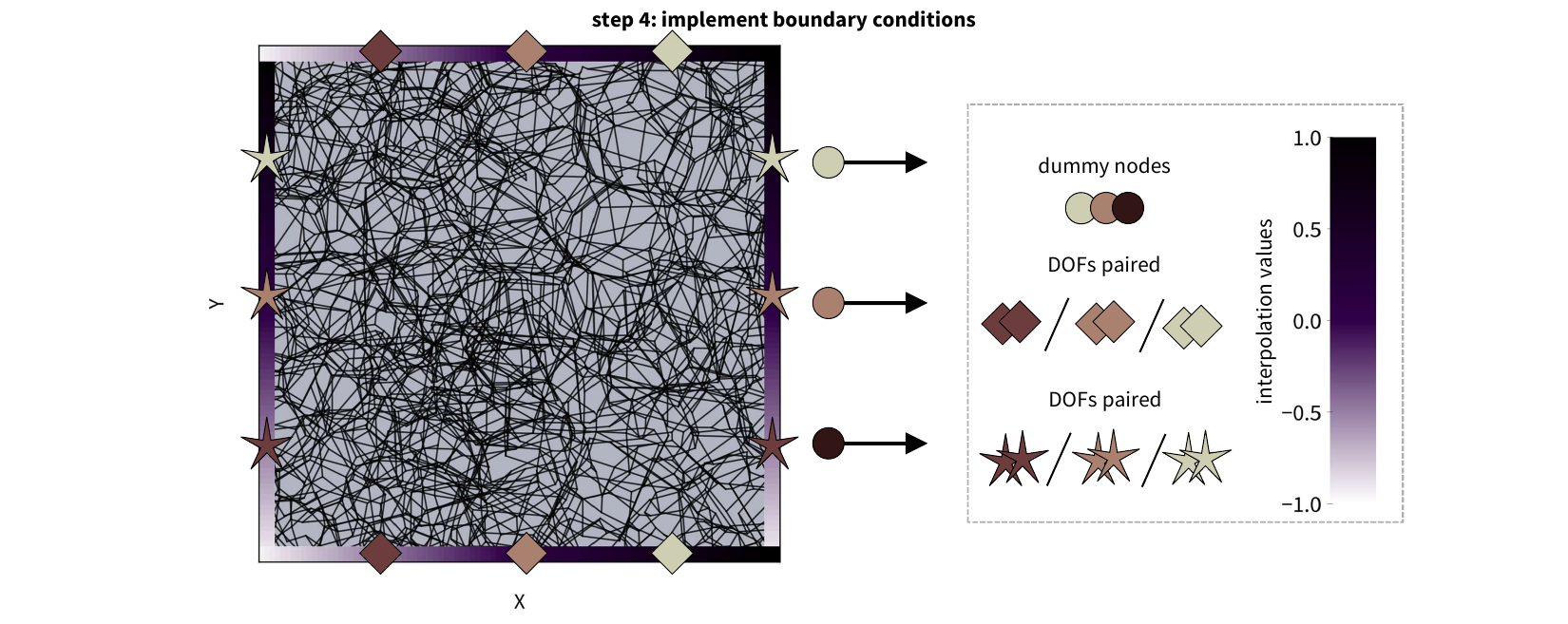}
\caption{\textbf{Pipeline for generating periodic topology and applying periodic boundary conditions.} 
To achieve a periodic distribution of the boundary nodes, we tile the initial seeds in 26 replicas around the central domain (step 1) and tessellate the 3D space with a Voronoi diagram (step 2). 
Consequently, we map the nodes at the boundaries with their periodic counterpart (light purple dashed lines) and store the pairs to define the PBCs (step 3). 
Finally, we constraint the motion of node pairs located on opposing faces and ensure volume-preserving conditions by using an interpolation map on the free edges (step 4). 
For illustrative purposes, we show only tensile loading and implemented the conditions on the 2D case.}
\label{fig: PBC} 
\end{figure}

Figure \ref{fig: PBC} highlights how we enforce periodicity at the boundary nodes while maintaining the internal microfluctuations. 
First, we randomly seed our RVE domain with N seeding points and generate 26 periodic replicas of this domain by offsetting the seeding points around the central domain (Figure \ref{fig: PBC}, step 1).  
Next, we apply Voronoi tessellation using 27 X N points as seeding particles and extract the interconnected fibers from the skeletonized diagrams (Figure \ref{fig: PBC}, step 2). 
We crop the larger network to the original cubic domain and assemble a periodic edge connectivity map between each boundary node and its periodic counterpart on the opposing edge (Figure \ref{fig: PBC}, step 3). 
Finally, we implement PBCs by constraining the relative displacements of all node pairs, as defined in our assembled periodic boundary node connectivity table, to individual dummy nodes. 
These dummy nodes, unattached to any fiber in the model, serve as reference points used in the nodal equations that impose the periodicity of the nodally paired loading conditions \citep{wu2014applying}. 
In particular, at each individual dummy node, we enforce an affine macroscopic deformation gradient corresponding to the macroscopic RVE face deformation profile under study (e.g., uniaxial extension, biaxial extension, or shear, as illustrated in Figure \ref{fig: PBC}, step 4). 
For each of these macroscopic deformation profiles, we introduce a linear interpolation map to constrain the boundary nodes to remain planar during shear or uni/biaxial tests.

Upon solving the mechanical equilibrium of our topologically optimized RVEs under various periodic loading conditions, we obtain the reaction forces on our PBC-constrained faces. 
We use these reaction forces to compute a representative stress tensor across the discrete fiber network by calculating the first Piola–Kirchhoff stress tensor ($\mathbf{P}$):
\begin{equation}
   \mathbf{P} = \frac{1}{A_0} \sum_{i=1}^{N_i}  \mathbf{f}_i.
\label{eq: FPK_Stress} 
\end{equation}
where $A_0$ denotes the initial surface of our cubic RVE, ${N_i}$ are the dummy nodes, and $\mathbf{f}_i$ represents the resulting external forces acting on these nodes.
We account for geometrical nonlinearity throughout the solution of these boundary value problems.

In our mechanical model, pipeline-generated node locations and their interconnections form the geometric input for finite element analysis. Here, one element represents a single fiber.
Focusing purely on elastic effects, we assume fiber connections to be permanent, with no cross-link formation or branch breakage.
We also consider that the persistence length of fibers, defined as the characteristic length along the polymer chain over which its direction remains correlated before bending, is typically larger than the contour length.
As a result, the fibers are bending-stiff, and their thermal fluctuations can be neglected \citep{Burla2019a}.
While the Euler-Bernoulli beam formulation could have been considered \citep{Shahsavari2012}, neglecting shear stiffness is appropriate for fibers that are highly slender, with a cross-section-to-length ratio of less than 1:15.
However, experiments in biological networks (e.g., collagen and fibrin) reveal a log-normal distribution of fiber lengths, which results in short fibers that violate this requirement \citep{Lindstroem2010, Lindstroem2013}. 
Therefore, we model individual fiber mechanics using second-order Timoshenko shear-flexible beams \citep{Peirlinck2017}, which are particularly suited for representing biological networks \citep{Shahsavari2013}.
We assume our fibers behave linearly elastic in both extension and compression, while accounting for fiber softening under compression.
This asymmetry in the elastic modulus reflects the well-established observation that biological fibers soften their mechanical response under compression \citep{Notbohm2015, Janmey2006}. 
Following the approach of \cite{Liang2016}, we implement this softening through a constitutive law applied to bending-resistant elements, while maintaining a linear elastic response in tension. 
As a result, our final model is bilinear, with the compressive elastic modulus set to one-tenth of the tensile modulus, consistent with \citep{Notbohm2015}.
To incorporate this bilinear tension-compression asymmetry, we developed a custom user-defined field USDFLD subroutine that defines the fiber elastic modulus at each integration point based on the axial strain of the loaded beam. 

\subsection{Case study: rheological tests on collagen networks}
\label{Collagen}
We evaluate the role of microstructural features, including average connectivity, average fiber length and thickness, fiber elastic modulus, and network concentration, on the differential elastic modulus, defined as the derivative of the stress with respect to the shear strain, as in \cite{Jansen2018}.
Our analysis employs the first Piola–Kirchhoff stress measure $\mathbf{P}$ (Equation \ref{eq: FPK_Stress}), which is first smoothed and then differentiated with respect to RVE shear to yield a nominal differential shear modulus.
We compare our results with experimental data taken from \cite{Jansen2018}, where the authors investigated (i) the nonlinear elastic behavior of rat tail collagen type I networks polymerized at different temperatures using a rheometer and (ii) network architecture features, including connectivity, fibril diameter, and length through multiple imaging modalities. 
Briefly, Jansen and coauthors found that higher polymerization temperatures reduce the average connectivity and the fibril diameter (Table \ref{tab:CollagenStructure}). 
We virtually replicate these polymerized networks and their respective microstructural architectures by matching the in silico collagen concentration ($\rho_c$) and the experimental values, as in \cite{Eichinger2021}, through:
\begin{equation}
    \rho_c = \frac{L_{tot} \pi R_f^2 }{V_{RVE} v_c}.
\label{eq: concentration}
\end{equation}
where $R_f$ is the fiber radius, $L_{tot}$ represents the sum of all individual fiber lengths, $V_{RVE}$ denotes the volume of the RVE, and $v_c$ is the specific volume of collagen fibers \citep{Hulmes1979}.

We create 40 x 40 x 40 \textmu$\mathrm{m^3}$ RVEs in which we model collagen fibrils as circular Timoshenko beams which - based on the respective polymerization temperature we aim to replicate - have diameters ranging from 160 nm to 300 nm and average lengths ranging from 1.6 to 3 \textmu m (Table \ref{tab:CollagenStructure}).
We keep connectivity constant while varying the number of seeds within the fixed domain size and set the fiber's tensile modulus equal to 700 MPa (\cite{Gacek2023}). 
We verify that these alterations consistently maintain the fiber radius-to-length ratio around $10^{-3}$ (\cite{Licup2015}).
We assume that concentration remains constant during polymerization. 


\section{Results}
\label{Res}

\subsection{Generation of microstructure-informed fiber networks}
\label{Gen}
\begin{figure}[H]
\includegraphics[width=\textwidth]{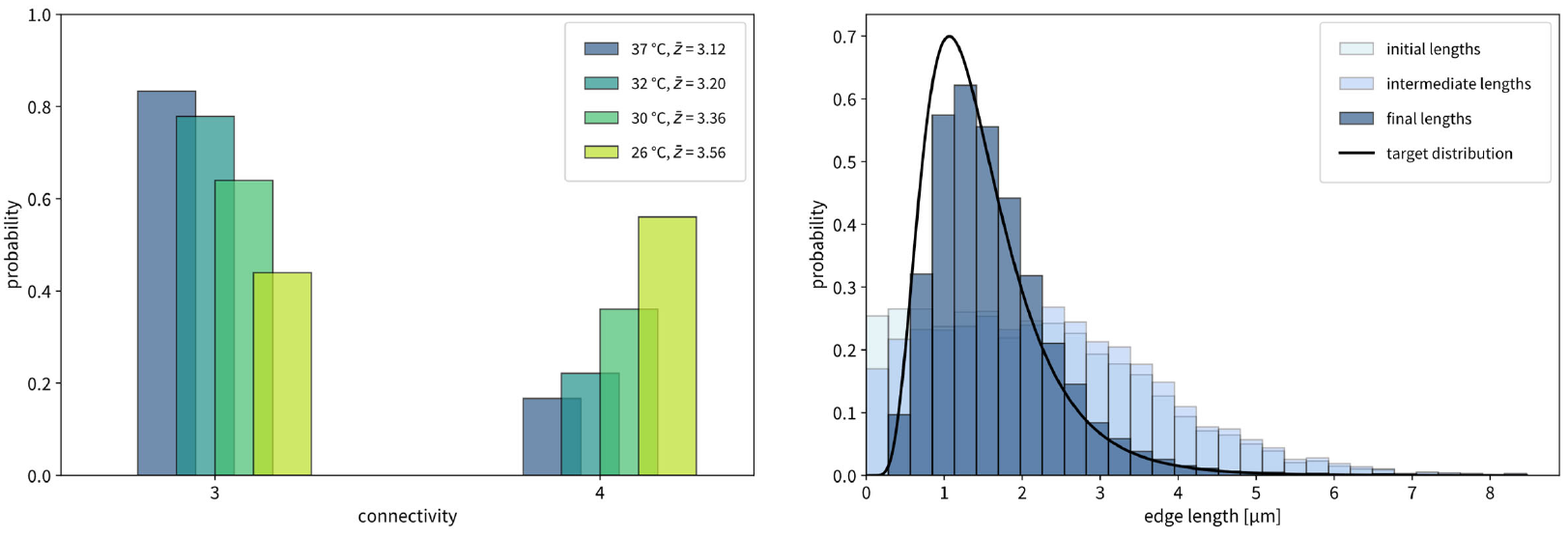}
\includegraphics[width=\textwidth]{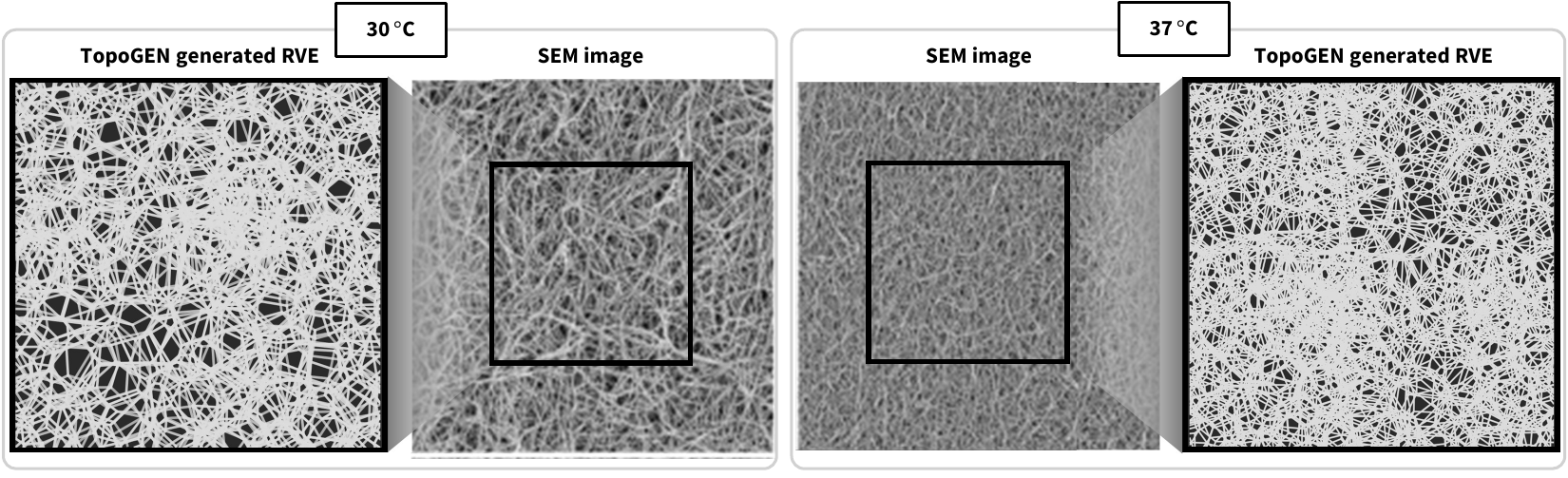}
\caption{\textbf{TopoGEN in action, experimentally informed discrete fiber network topology optimization.} 
The top row shows two-step connectivity and length optimization results across temperatures.
Increasing the temperature decreases the network's average connectivity ($\overline{z}$), leading to more branches forming between fibers (left plot).
Weighted edge removal facilitates an intermediate optimization stage, where shorter edges are preferentially removed during system dilution. 
Subsequently, the algorithm shifts the intermediate state toward the final optimized configuration by rearranging the network's nodes. 
The length optimization is shown here for the 37~\textdegree{}C condition at a concentration of 4 mg/mL  (central plot).
The four images on the bottom row compare scanning electron microscopy (SEM) images \textit{(taken from \cite{Jansen2018})} of reconstituted collagen I network in vitro (central images) with the simulated maximum intensity projection of our three-dimensional in silico reconstructions (external images) at 30~\textdegree{}C and 37~\textdegree{}C, respectively. 
The black squares in the experimental and simulation images have an edge length of 20 \textmu m.}
\label{fig: SimulatedAnnealing_Result} 
\end{figure}

Figure \ref{fig: SimulatedAnnealing_Result} highlights the power of our TopoGEN pipeline to generate various topology-informed discrete fiber networks. 

Without loss of generality, we here leverage TopoGEN to generate collagen type I hydrogels representing the experimental in vitro networks generated by \cite{Jansen2018}.
Starting from random network initializations, we use simulated annealing to align the network architecture in terms of connectivity and edge length distributions, prioritized as key parameters for capturing isotropic network properties.
To match the connectivity observed in self-assembled collagen type-I networks, where fibers form branches (connectivity 3) or cross-links (connectivity 4), we constrain our network to accept only these connectivity values.
In particular, our generated discrete fiber network topologies replicate the observed decrease in average cross-link density associated with increasing temperatures.
Therefore, as the temperature rises, inter-fiber interactions shift towards a prevalence of branch points over cross-links, as can be seen from the decreasing connectivity in the top-left subplot of Figure \ref{fig: SimulatedAnnealing_Result}. 
The length optimization follows a two-step process. 
First, we remove small, unrepresentative fiber segments during the connectivity optimization stage to ensure a physically meaningful network structure.
To achieve this, we assign higher removal probabilities to shorter edges and iteratively optimize connectivity following this weighted edge removal approach.
Next, we adjust node positions to align with the target length log-normal distribution using simulated annealing guided by the KL divergence metric.
This metric demonstrates superior performance, achieving more significant improvement within runtimes comparable to a conventional technique, such as the Cramér-von Mises (CVM) test (\cite{Lindstroem2010, Lindstroem2013, Eichinger2021}). 
See Table \ref{tab:CVM_VS_KL} for a detailed comparison across 20 independent tests of the performance metrics between the two methods.
The average fiber length is optimized to match the experimentally measured fiber lengths observed across different temperatures (Table \ref{tab:CollagenStructure}), and the resulting log-normal distribution is depicted in the top-right subplot of Figure \ref{fig: SimulatedAnnealing_Result}. 
In the bottom row of Figure \ref{fig: SimulatedAnnealing_Result}, two-dimensional projections of our three-dimensional in silico reconstructions are compared to high-resolution images of reconstituted collagen I network, taken from \cite{Jansen2018}.
For the illustrated topology, the optimization convergence is achieved within $10^5$ number of iterations in 81 seconds on a 64-bit system with a 12th Gen Intel(R) Core(TM) i9-12900H @ 2.50 GHz and 32.0 GB RAM. See Figure \ref{fig: performance} for details on the optimization performance.

\subsection{Boundary value problem at the mesoscale}
\label{PBC}
\begin{figure}[H]
\includegraphics[width=\textwidth]{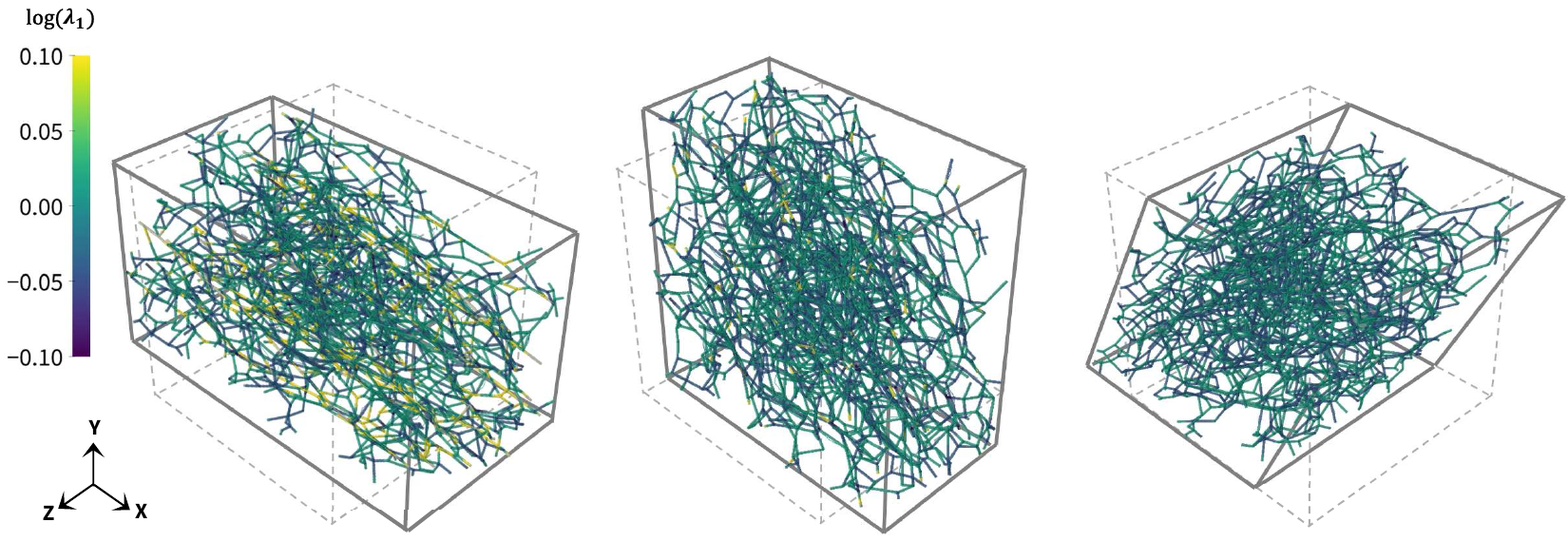}
\caption{\textbf{Multi-axial loading with periodic boundary conditions.} 
Fiber logarithmic stretch distribution in a representative (37~\textdegree{}C, average connectivity 3.12, concentration 4 mg/mL, average fiber length 2 \textmu m) RVE under uniaxial (left), biaxial (middle), and simple shear (right) loading, with 50\% stretch applied for each loading condition. 
The undeformed cubic domain is shown with dashed lines and the deformed domain with solid lines.} 
\label{fig: PBC_Results} 
\end{figure}

Figure \ref{fig: PBC_Results} showcases an exemplary finite element simulation outcome of a representative fiber network generated at 37~\textdegree{}C, with an average connectivity of 3.12, a concentration of 4 mg/mL, and an average fiber length of 2 \textmu m. 

The generated RVE undergoes uniaxial, biaxial, and simple shear deformations with fully periodic boundary conditions. 
We conduct all our RVE mechanical deformation analyses using Abaqus/Standard. 
The average number of degrees of freedom in our simulations is \( 5 \times 10^{4} \).
To ensure convergence, we adhere to Abaqus' default convergence criteria and employ automatic stabilization. We strictly require that at every increment, the ratio of viscous damping energy (reported as ALLSD) to total strain energy (reported as ALLSE) does not exceed 2\% \citep{Peirlinck2017,Kakaletsis2023}.

\subsection{Microstructure-function relationships in simulated collagen networks}
\label{Micro-Function}
To evaluate how each microstructural feature influences the network's differential shear modulus, we systematically vary each structural parameter based on experimental observations for collagen type I networks (Table \ref{tab:CollagenStructure}) and perform simple shear tests up to 50\% shear on ten representative samples per tested topological or microstructural alteration.
We analyze the impact that these topological and microstructural changes have on the differential shear modulus across increasing stretch levels relative to the average value of the original network architecture.

\begin{figure}[H]
\includegraphics[width=\textwidth]{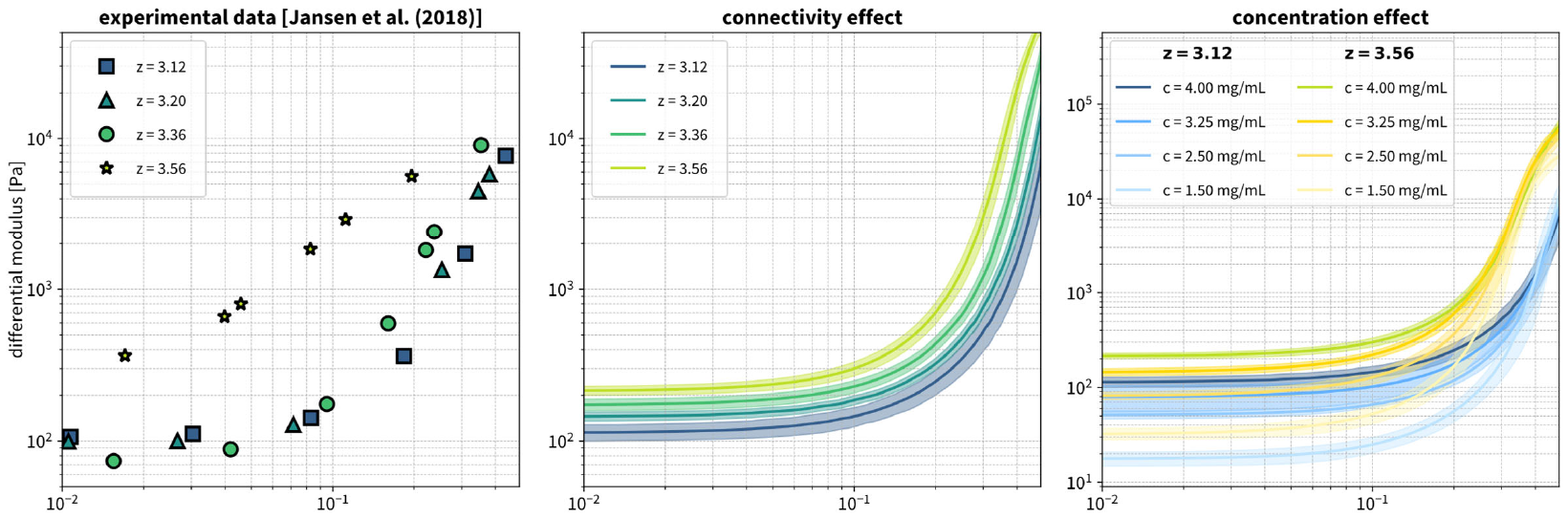}
\includegraphics[width=\textwidth]{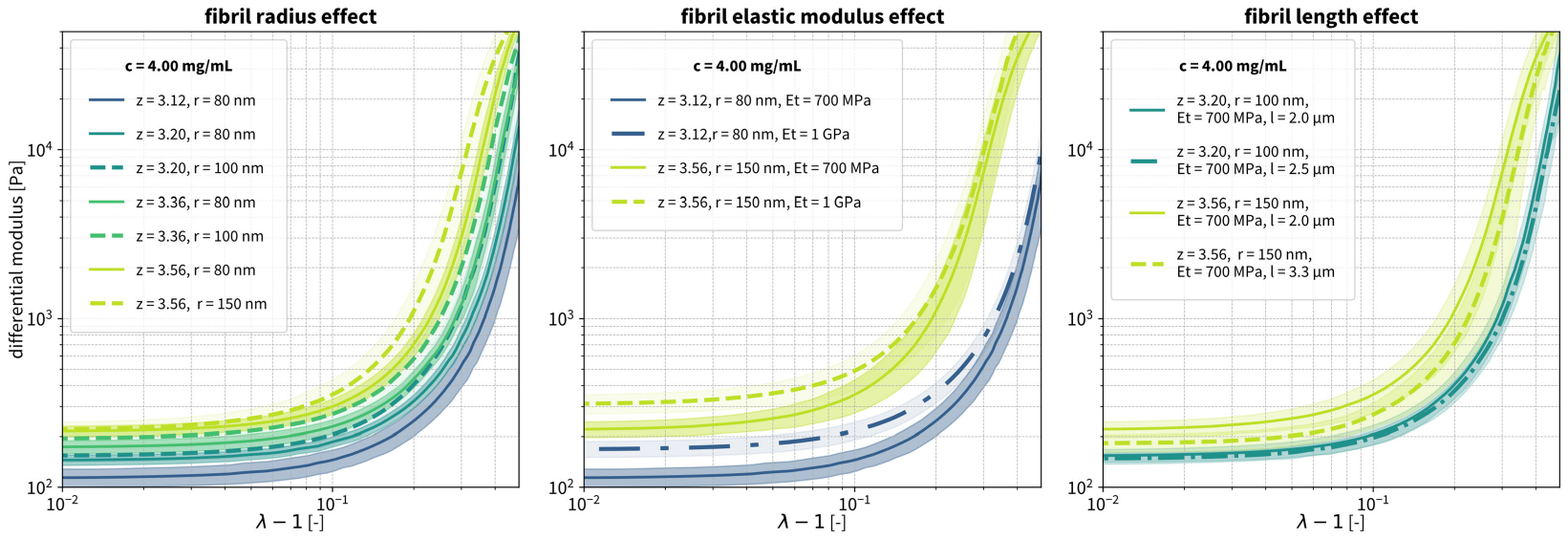}
\caption{\textbf{Analysis of microstructure-mechanics relationship in reconstituted collagen networks.} 
The differential shear modulus is shown as a function of rescaled stretch $(\lambda - 1)$ on a log-log scale for each tested condition.
The top row presents (1) experimental results from rheological tests on reconstituted collagen networks with a concentration of 4 mg/mL at varying average connectivity \citep{Jansen2018} and our simulation results from shear tests (up to 50\% shear) for (2) 4 mg/mL samples at the same connectivity tested experimentally and (3) decreasing concentrations at fixed connectivity of 3.12 and 3.56. 
In all the top row simulations, the fibril-related parameters are set to their minimum values, with fibril radius of 80 nm, elastic modulus of 700 MPa in tension and 70 MPa in compression, and an average length of 2 \textmu m.
The bottom row shows simulation results under the same loading conditions as the top row, focusing on the impact of collagen fibril morphology and elastic modulus. 
In particular, to study the fibril elastic modulus effect we increase the tensile and compressive elastic moduli to 1 GPa and 100 MPa, respectively, maintaining a compression-to-tension ratio of one-tenth as in \cite{Notbohm2015, Liang2016}.
These results examine inter- and intra-connectivity effects at a 4 mg/mL fixed concentration. 
All simulation results represent the response of ten samples per condition, with the average response shown as a line and the standard deviation indicated by the shaded area.}
\label{fig: CollaGEN} 
\end{figure}

Figure \ref{fig: CollaGEN} illustrates the mechanical response for various topological and microstructural parameters derived from experiments and our simulations.  
We examine the role of network-level factors (i.e., connectivity and concentration) and fibril-level characteristics (i.e., radius, elastic modulus, and average length) using our 3D simulations. 
We systematically alter these parameters within the experimentally reported ranges from \cite{Jansen2018} (Table \ref{tab:CollagenStructure}) for diameter and average fiber length, and from  \cite{Mahutga2023, Gacek2023} for fibril elastic modulus.
We find that transitioning from the low-stretch to the high-stretch regime results in a nonlinear stiffening for all tested topologies.
In particular, for all collagen type I networks, we observe a two-order increase in stretch stiffening from the initial to the maximum differential shear modulus, with significant stiffening emerging in the medium-to-high-stretch regime.
Our model results are consistent with the experimental and computational data from \cite{Jansen2018}, where the concentration is fixed at 4 mg/mL while the average connectivity varies with temperature (Figure \ref{fig: CollaGEN} - experimental data and connectivity effect). 
Conversely, variations in concentration exhibit an uneven impact, with the most significant changes in the differential shear modulus occurring in the low-stretch regime (Figure \ref{fig: CollaGEN} - concentration effect). 
In this regime, fibrils primarily undergo non-affine rotations to align with the principal loading direction.
Notably, increasing the concentration from 1.5 to 4 mg/mL results in an order-of-magnitude increase in the initial differential shear modulus.
In contrast, within the large-stretch regime, the maximum modulus remains largely insensitive to concentration changes. 
At a concentration of 1.5 mg/mL, our models predict an almost three-order-of-magnitude increase in modulus when transitioning from the low- to high-stretch regime for both tested topologies.
These results suggest that, in the nonlinear regime, network stiffness becomes independent of fiber concentration, consistent with the findings of \cite{Sharma2016, Licup2015}.
With regard to the morphological changes of the fibers, variations in radius and length made to simulate the temperature do not exhibit noticeable effects across all tested topologies and stretch ranges (Figure \ref{fig: CollaGEN} - fibril radius and length effects).
Overall, increasing the radius effect slightly increases the network's differential shear modulus, whereas increasing the length reduces it. 
With the more pronounced effect in the low-stretch regime.
This is because longer, more slender fibrils are more susceptible to bending, particularly in the low-stretch regime when not aligned with the primary loading direction.
The limited impact arises from the narrow range of morphological variations tested, which are based on experimentally observed temperature-induced changes. 
These variations preserve a nearly constant bending-to-stretching stiffness ratio, therefore preventing significant changes in the network’s mechanical behavior.
Increasing elastic modulus from 700 MPa to 1 GPa (Figure \ref{fig: CollaGEN} - fibril elastic modulus effects) yields comparable behavior for the medium-to-high-stretch regime. 
However, it exerts a more pronounced influence in the low-stretch regime. 
Notably, these findings remain consistent across different connectivity conditions, suggesting that they are independent of the topology characterizing the network.
Therefore, including the asymmetry in the fibril elastic modulus introduces an uneven effect in the nonlinear response of our networks.
In contrast, when tension-compression asymmetry is excluded, the fiber elastic modulus directly scales with the network’s differential modulus (Figure \ref{fig: unimod}), and nonlinearity primarily arises from geometry.

\begin{figure}[H]
\includegraphics[width=\textwidth]{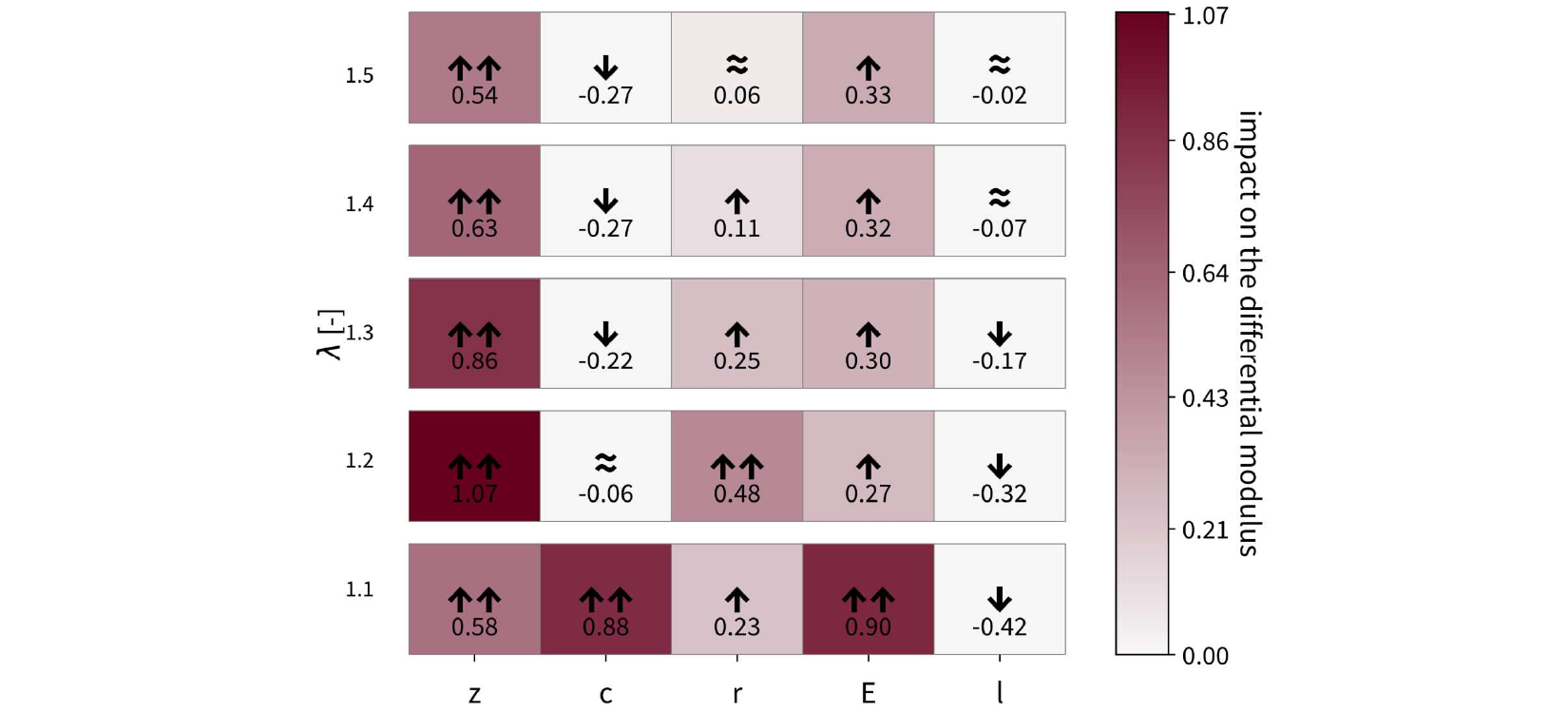}
\caption{\textbf{Microstructure effect on differential shear modulus across different stretch ranges.} 
Each cell shows the relative impact score (difference in mean modulus normalized by the combined sum of their averages and the pooled standard deviation) for a given parameter at a specific stretch $(\lambda)$.
Arrows indicate qualitative impact strength ($\approx$: negligible, $\uparrow$ / $\downarrow$: moderate, $\uparrow\uparrow$ / $\downarrow\downarrow$: strong), while the numbers below represent the exact relative impact values.
Parameter labels: $z$ for average connectivity, $c$ for concentration, $r$ for fibril radius, $E$ for fibril elastic modulus, $l$ for fibril average length.}
\label{fig: DiffEffect} 
\end{figure}

Figure \ref{fig: DiffEffect} summarizes the deformation-dependent influence of each microstructural parameter on the network’s bulk elastic properties.
To build the central table, we compare the variation in the differential shear modulus for each microstructural parameter by averaging data from 10 independent samples for each tested condition.
At predetermined stretch points, we compute the mean and standard deviation for each condition and derive the difference between the means normalized by the combined sum of their averages and the pooled standard deviation.
We aggregate these normalized changes across all parameters and compute the 40\textsuperscript{th} and 70\textsuperscript{th} percentiles to set thresholds: values below the 40\textsuperscript{th} percentile indicate a low impact, values between the 40\textsuperscript{th} and 70\textsuperscript{th} percentiles indicate a moderate impact, and values above the 70\textsuperscript{th} percentile indicate a high impact. 

\subsection{Non-affinity in simulated collagen network}
\label{Non-affinity}
\begin{figure}[H]
\includegraphics[width=\textwidth]{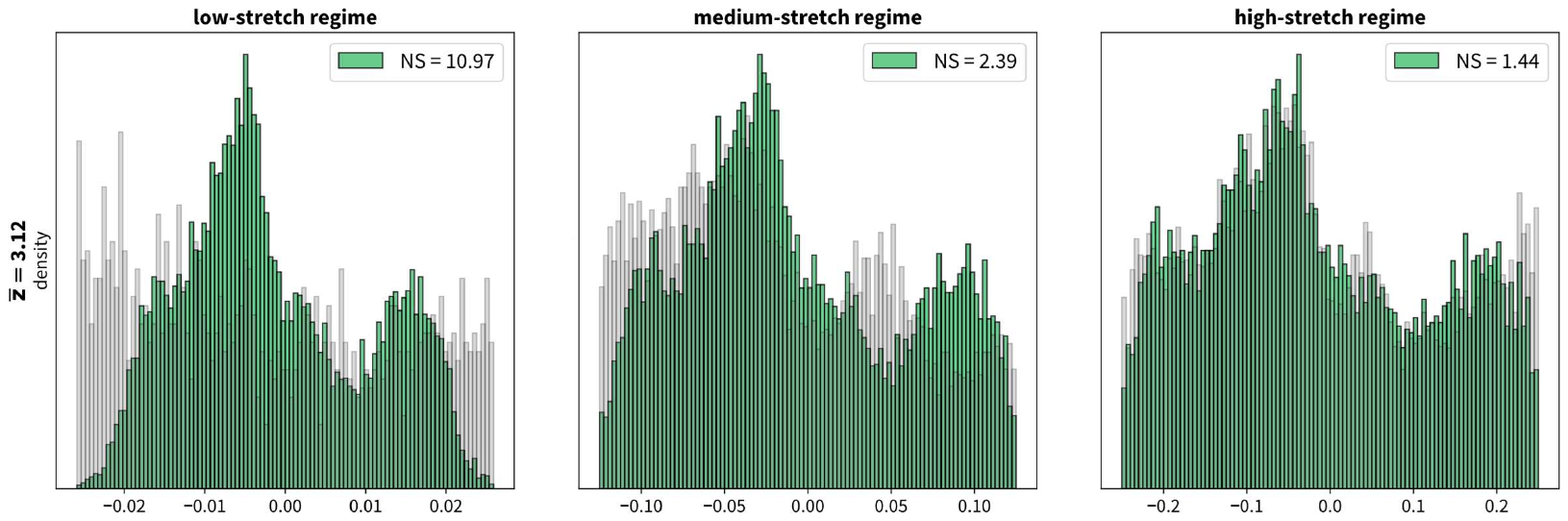}
\includegraphics[width=\textwidth]{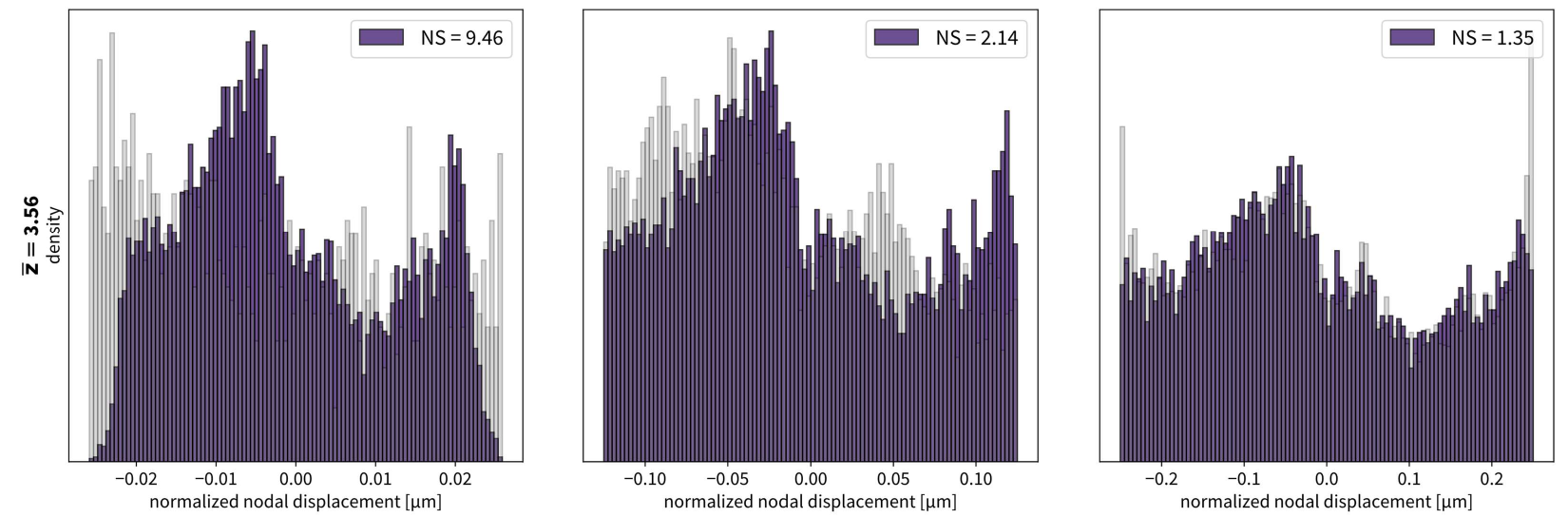}
\caption{\textbf{Non-affinity score with nodal displacements.} 
Comparison of predicted affine displacements (gray) and computed displacements for five samples, with average connectivity $\overline{z}=$ 3.12 (top row) and connectivity $\overline{z}=$ 3.56 (bottom row), at three distinct stretch levels. 
For the low-stretch regime, non-affine modes predominate, with a large set of nodes not following the boundary displacement based on the macroscopic deformation gradient. 
At stretch 1.5, stretch-induced affinity enhances the agreement between predicted and measured distributions.} 
\label{fig: NonAffineScore} 
\end{figure}

Figure \ref{fig: NonAffineScore} illustrates the discrepancy between the nodal displacements predicted by our in silico simulations and the affine predictions based on the deformation gradient imposed at the boundaries.

Here, we focus on two specific cases: loosely connected networks with an average connectivity of $\overline{z}$ = 3.12, and more interconnected networks with average connectivity of $\overline{z}$ = 3.56. 
We define the non-affinity score ($NS$) as the normalized root mean squared error between purely affine deformation ($u_{affine}$) of the whole network (i.e., the displacements obtained by interpolating the macroscopic deformation gradient imposed at the boundaries to all internal nodes)  and the displacements predicted in our simulations ($u_{predicted}$):
\begin{equation}
NS = \sqrt{\frac{\displaystyle\sum_{i=1}^{n}\Bigl(u_{predicted}^{(i)} - u_{affine}^{(i)}\Bigr)^2}{\displaystyle\sum_{i=1}^{n}\Bigl(u_{predicted}^{(i)} - \bar{u}_{predicted}\Bigr)^2}}
\label{eq: NS}
\end{equation}
where $n$ is the total number of nodes.
The networks exhibit different behavior from the affine model under all the tested conditions. 
The most significant discrepancies occur in the low- to medium-stretch regime, where the $NS$ score decreases fivefold. 
In contrast, the medium- to high-stretch regime experiences only a twofold reduction. 
In the low-stretch regime, very few nodes adhere to the boundary displacement prescribed by the imposed macroscopic deformation gradient. 
However, a noticeable shift occurs in the medium- to high-stretch regime, where nodal displacements closely follow an overall affine displacement pattern.
Both loosely connected (with $\overline{z}=$ 3.12) and more interconnected (with $\overline{z}=$ 3.56) networks converge toward a more affine response at higher stretch levels.
Networks with lower connectivity exhibit slightly increased non-affinity in the low-stretch regime than those with higher connectivity.
With fewer fibers aligning with the affine deformation, load distribution in these loosely connected networks becomes highly localized. 
Increasing the number of cross-links compared to the branches (with $\overline{z}=$ 3.56) results in a marginally improved match between the theoretical affine predictions and the experimental measurements.
Overall, these networks exhibit nodal deformation patterns that, in the low- to medium-stretch regime, deviate from affine deformation assumptions.

\section{Discussion}
\subsection{Topological complexity and heterogeneity }
Many materials - not just biological materials - are made of fibers. 
Natural fibrous materials include cotton, wool, and silk, while industrial examples range from polymeric foams and rubber to felt and paper.
Among biological materials, an important fibrous system is the extracellular matrix, where collagen fibers assemble into networks to bear mechanical loads and transmit mechanical information over long distances \citep{Ban2016}. 
Another crucial fibrous system is fibrin, derived from fibrinogen, which assembles into a fibrillar network upon tissue injury to form blood clots. These clots must withstand high tensile forces from blood flow while ensuring rapid degradation after wound healing to prevent thrombosis \citep{Pieters2019}.
Accurately modeling this seemingly passive fibrous environment through computational methods remains a significant challenge. 
A key requirement is to account for the three-dimensional arrangement of these networks and their structural heterogeneity, which affects the local cellular environment and influences the overall mechanical response of tissues.
In this study, we employ a mechanistic approach that explicitly models the interconnected fiber networks underlying soft tissues by leveraging high-resolution imaging data. 
This framework enables us to investigate how macroscopic material behavior emerges from its microscopic constituents, accounting for microstructural variations.
Our generative pipeline builds on the framework by \cite{Lindstroem2010}, where high-resolution confocal images are adopted to derive an image-based representation of collagen networks using simulated annealing.
The iterative optimization process is designed to transform initially random Voronoi networks into structures that accurately represent the target biological network in terms of connectivity and length distribution. 
Compared to \cite{Lindstroem2010}, we capture the non-affine localized effects by restricting the network to include only cross-links and branches.
For the length optimization, we employ Kullback-Leibler divergence to compare the actual and target distributions throughout the iterative optimization process. 
As a result, we improve the energy accuracy of the optimization process without increasing runtimes compared to the commonly used \citep{Lindstroem2010, Lindstroem2013, Nan2018, Eichinger2021} Cramér-von Mises tests. 
Notably, while the probabilistic search strategy of simulated annealing cannot guarantee convergence of the optimization scheme to the global minimum \citep{Suman2006}, it is sufficient for our purposes that the final configuration statistically aligns with experimental measurements without requiring exact correspondence. 
While focused on collagen-like networks, our modeling pipeline can adapt to other biomaterials and topologies. 
For instance, we can effectively tune the fibers to replicate the morphology and mechanics of fibrin networks to simulate the mechanical behavior of another essential load-bearing biopolymer in the extracellular space. 
Alternatively, we can introduce higher connectivity by merging nodes within a specific distance threshold, or we can disregard the bending rigidity to model fibers as springs transmitting only central forces, as observed in the intracellular environment of semi-flexible F-actin networks and intermediate filaments \citep{Licup2015}.  
Leveraging detailed microstructural information, our approach can be extended to model non-living networks. 
For example, with random Voronoi seeds, we can replicate the microstructure of paper, while with regularly arranged seeds, we can create ordered structures, such as those found in fabrics \citep{Picu2011}.


\subsection{Impact of microstructure on stress-stretch behavior, a case study of collagen networks}
To assess the robustness of our generative pipeline, we subject a library of topologically rich RVEs to shear tests. 
In constructing the library, we incorporate the experimental evaluations reported by \cite{Jansen2018}, wherein temperature variations induce a change in average connectivity by altering the number of cross-links and branches. 
Beyond influencing connectivity, temperature variations also induce morphological changes in the fibers, affecting their average length and diameter.
We account for all such microstructural variations and summarize our findings in Figure \ref{fig: DiffEffect}.
Consistent with trends reported in the literature, our results confirm that each microstructural parameter has a distinct influence on the differential modulus \citep{Jansen2018, Licup2015, Sharma2016}.
Specifically, our results reveal a uniform influence of connectivity across the stretch range, in agreement with prior work \citep{Jansen2018} identifying local connectivity as a key determinant of collagen network elasticity. 
In contrast, collagen concentration predominantly affects the low-stretch regime.
This sensitivity of the low-stretch mechanical response to collagen concentration is aligned with prior theoretical predictions: \cite{Licup2015} showed that, in this regime, the elastic energy is dominated by bending modes, and the linear shear modulus scales quadratically with the fiber volume fraction.
The nonlinear stiffening, however, becomes essentially insensitive to network concentration at higher deformations, a behavior that appears to be unique to collagen \citep{Licup2015} and fibrin networks \citep{Jansen2013}.
To verify that the effect of collagen concentration on the low-stretch mechanical response is independent of network connectivity, we examine the differential modulus across varying concentrations and connectivity levels, consistently observing the same concentration-driven trends.
Changes in fibril radius and length do not produce large shifts in the differential shear modulus, largely due to the limited parameter range explored, deliberately constrained to match experimental observations. 
Within this range, the bending-to-stretching stiffness ratio remains relatively constant, limiting the impact on network mechanics across stretch regimes. 
Nevertheless, even modest slenderness variations can lead to measurable stiffness changes, as network mechanics in the bending-dominated low-stretch regime is highly sensitive to fiber length and radius, in line with prior modeling studies \citep{Broedersz2012, Islam2018, Sharma2016}.
In contrast, the intrinsic elastic modulus of the fibrils plays a significant role in modulating the variation of the differential modulus, especially in the low-stretch regime.
The uneven low-to-high stretch regime effect of the fibril elastic modulus on the differential modulus - consistently observed across different polymerization conditions and network topologies - emerges from the imposed tension-compression fibril modulus asymmetry.
This is supported by our simulations, where removing this intrinsic fiber elastic modulus asymmetry eliminates the effect.
These results are important because they allow us to unravel the impact of various microstructural features that are influenced by environmental features such as temperature, concentration, and polymerization, highlighted by \cite{Andriotis2023} to have a crucial impact on the behavior of the collagen network.
Indeed, while previous studies have focused on the extraction of network parameters from gels polymerized at different concentrations \citep{Lindstroem2010, Lindstroem2013}, experiments show that concentration alone cannot fully characterize the network structure (i.e., mean connectivity) and the mechanisms of fibril formation (i.e., fibril length, radius, and elastic modulus).
Our work presents a mechanistic approach for modeling discrete fiber networks that explicitly incorporates microstructural variability. 
This method is particularly valuable for informing experimental investigations, where the challenge lies in deciphering how individual microstructural features influence the overall mechanical response. 
This complexity arises from the interplay of multiple multiscale factors, which are difficult to disentangle using typical laboratory protocols.
Moreover, because microstructural effects can evolve during rheological loading, typically manifesting as a nonlinear increase in global network stiffness, any comprehensive analysis must account for stretch-dependent behavior. 
Our models thus offer a powerful tool to support experimental assessments of these intricate interactions.
In our case study, we focused on the influence of concentration and temperature, as reported by \cite{Jansen2018}, but in the future, we could think of expanding the library by looking at the combined effect of these two parameters or the effect of new environmental parameters, such as pH.
For assessing the topology and microstructure effect on differential shear modulus across different shear ranges, we assume independence among all the structural parameters. 
We must underline that this assumption introduces limitations because we disregard the potential existence of confounding factors. 
Future studies need to systematically isolate confounding factors when analyzing the nonlinear elastic response of collagen or other types of biopolymer networks. 
Another assumption in our work is that concentration remains constant during polymerization. 
Advanced imaging could test whether temperature-induced affects fiber volume fraction and provide a clearer understanding of how concentration changes during polymerization.
In this work, we focus exclusively on elastic effects, assuming that fiber connections remain permanent with no cross-link breakage. 
To further enhance the microstructural analysis, future studies should consider time-dependent effects across varying polymerization conditions, including the irreversibilities introduced by cross-link breakage.

\subsection{Impact of microstructure on non-affinity}
To simulate biopolymer networks polymerized under varying conditions, we optimize random networks with data derived from high-resolution images. 
While previous studies \citep{Lindstroem2010, Lindstroem2013, Nan2018, Eichinger2021, Mahutga2023, Leng2021} treated networks with multiple fibers intersecting at a single node, our model takes a different approach. 
We exclusively consider fiber-fiber interactions as permanent cross-links or branches, deliberately avoiding higher-order connections that fail to replicate the loosely connected architecture characteristic of collagen networks.
This choice aligns with the concept of non-affinity that has been increasingly highlighted in recent experimental \citep{Cavinato2020} and computational \citep{Mahutga2023, Chen2023, Parvez2023} works as a crucial factor in understanding local alterations in the extracellular space. 
Non-affinity arises from the loose interconnections \citep{D’Amore2010, Burla2020} between fibers that, in the case of networks with only central-force interactions, fall below the Maxwell isostatic threshold (\cite{Maxwell1864}). 
According to the Maxwell criterion, a three-dimensional central force network is stable only if the average number of fibers per node exceeds six. 
In isotropic networks, where nodes are branches or cross-links, each branch corresponds to three crossing fibers, and each cross-link corresponds to four. 
This configuration results in an average number of fibers per node that falls below the Maxwell threshold. 
In such networks, if the fibers behave solely as axial load-bearing elements such as springs, or if bending modes are allowed within the fiber but not transmitted at the nodes by using pin-jointed cross-links \citep{Picu2011}, the structure is prone to collapse.
However, the structural integrity of networks of stiff filaments - such as collagen networks - is maintained either through intrinsic stabilizing effects of the fibers (e.g., bending rigidity) or extrinsic factors related to their surrounding environment (e.g., loading conditions). 
We reflect this principle in our computational model, where stability is ensured by (1) athermal collagen fibrils with a bending-to-stretching stiffness ratio on the order of $10^{-3}$, as demonstrated in \cite{Licup2015}, and (2) boundary-applied stretch, which stabilizes the system by transitioning it from a non-affine to an affine configuration where fibers align with the loading direction and deform consistently with the macroscopic deformation gradient. 
We define $NS$ as a metric to quantify the level of non-affinity by measuring the deviation of cross-link displacements obtained from simulations relative to those predicted under a purely affine deformation assumption.
The outcome of this score is consistent with previous studies \citep{Broedersz2011, Sharma2016}, which demonstrate that non-affinity decreases with increasing average connectivity and applied stretch.
Additionally, the inconsistency between microscopic and macroscopic affine displacements is most evident in the low-stretch regime for both tested topologies.
In this regime, theoretical models predict a more uniform load distribution across the network.
This observation confirms that biological networks adopt a highly heterogeneous load distribution strategy and align with previous works \citep{Chandran2005, Cavinato2020, Mahutga2023}.
This loading heterogeneity is most pronounced during the low-to-medium-stretch regime, where most of the fibers rotate or bend while just a few actually stretch.
With these findings, we highlight the importance of heterogeneous three-dimensional models in accurately capturing localized phenomena in the space surrounding cells and provide an in-depth insight into the mechanisms of gradual fiber recruitment.
Other non-affinity measurements have been proposed in the literature to quantify this important phenomenon from a computational perspective \citep{Picu2011, Shivers2023}.
In line with these methods, we compare the model-predicted displacements to those expected from a homogeneous continuum-scale affine deformation map.
Future research should focus on systematically extracting quantitative measures of non-affinity in vitro.


\subsection{Untangling the network puzzle}
We examine the influence of the microstructure on the shear differential modulus of collagen networks polymerized at various temperatures. 
Overall, for all the tested conditions, our analysis highlights a typical three-phase response in discrete fiber networks under load. 
In the linear elastic phase, the differential modulus remains constant as stretch increases, and geometric nonlinearity is negligible. 
This is followed by the nonlinear stiffening phase, where fibers reorient along the loading direction. 
A subset of fibers, initially oriented perpendicularly, undergoes significant bending. 
This geometrically induced nonlinearity happens when non-affine, bending-dominated deformation modes develop. 
Finally, the stress-stretch curve becomes linear in the affine-dominated phase. 
In this region, stress paths form through fiber chains that connect the boundary faces and limit further structural changes.
These results are consistent with well-established findings in the literature \citep{Broedersz2014, Licup2015, Picu2019}.
Furthermore, our results from the connectivity effect tests show trends consistent with in vitro rheological tests \citep{Jansen2018}, demonstrating a two-decade increase in stiffness from the initial to the final stretch. 
However, the connectivity effect alone does not fully account for the observed phenomena. 
Specifically, networks with higher connectivity exhibited stiffer responses at low stretches and a less pronounced nonlinear effect at higher stretches in vitro. 
Yet, in our computational models and those of \cite{Jansen2018}, higher connectivity networks exhibit nonlinear stiffening patterns that align more closely with those observed in lower connectivity networks. 
This suggests that nonlinear elastic behaviors in networks polymerized at different temperatures cannot be fully explained by connectivity alone.
Our results in Figure \ref{fig: CollaGEN} highlight the impact that fibril-related parameters such as length, fibril radius, and stiffness have.
Another study \citep{DavoodiKermani2021} explored the effect of varying morphological and graph descriptors on the linear elastic modulus of different networks. 
While providing valuable insights, it lacks direct analysis of biological networks under varying environmental conditions and shear loads throughout the highly nonlinear stress buildup observed in rheological experiments.
In our work, we focus solely on experimentally relevant microstructural variations (as summarized in Table \ref{tab:CollagenStructure}) and evaluate the differential modulus as a function of the shear.
We propose a novel approach to modeling biologically reconstituted networks that strategically incorporates the variations introduced by polymerization conditions during their reconstruction. 
In parallel, this approach can also address tissue-specific variations in microstructures observed in-vivo, where other environmental factors can lead to localized changes in the microstructure that result in different mechanical responses.
In both cases, it is essential to consider additional structural and topological features to untangle the network puzzle fully. 
We should direct significant efforts toward extracting detailed network topologies from high-resolution images captured under varying polymerization conditions. 
Doing so can enrich our models with robust statistical structural variations and provide deeper insights into the structure-to-mechanics relationship across multiple scales.

\section{Conclusion}
Our study provides a detailed analysis of the effects of topology and microstructure on the nonlinear differential shear modulus of biological networks. 
Without loss of generality for other biological and non-biological fiber networks, we use our developed framework to replicate the loosely interconnected microstructural architectures typical of collagen networks, enabling a quantitative evaluation of localized non-affine phenomena within these complex three-dimensional environments.
We achieve this condition by optimizing the average connectivity of in silico networks virtually replicating the temperature modification effect on the topology. 
Additionally, we expand the range of studied conditions by assessing other microstructural features beyond connectivity. 
Specifically, we demonstrate that even within a limited range of conditions - defined by structural variations in fiber morphology, their asymmetric response to tension-compression, and their concentration - these features influence network mechanics across different deformation regimes. 
By aligning our computational models with experimentally derived microstructural characteristics, we propose a generative pipeline that, while rooted in network theory, is foremost designed for experimental application.
This pipeline focuses solely on structural parameters that can be experimentally controlled and will benefit from further investigations into how topology changes under varying environmental conditions and, consequently, how these changes affect mechanical responses.

\section*{CRediT authorship contribution statement}
\textbf{Sara Cardona}: Writing – original draft, Methodology, Investigation, Visualization, Validation, Software, Data curation, Conceptualization
\textbf{Mathias Peirlinck and Behrooz Fereidoonnezhad}: Writing – review \& editing, Resources, Methodology, Investigation, Data curation, Conceptualization.

\section*{Declaration of competing interest}
The authors declare that they have no known competing financial interests or personal relationships that could have appeared to influence the work reported in this paper.

\section*{Acknowledgement}
The authors thank F. Gijsen and G. Koenderink for useful discussions.
This work was supported by the Delft University of Technology Startup Grant awarded to MP and BF, and the NWO Veni Talent Award 20058 to MP. 

\bibliographystyle{elsarticle-harv} 
\bibliography{references}  

\section{Supplemental Information}
\setcounter{figure}{0}
\setcounter{table}{0}
\makeatletter
\renewcommand{\thefigure}{S.\arabic{figure}} 
\renewcommand{\thetable}{S.\arabic{table}}  
\renewcommand{\fnum@figure}{Figure~\thefigure} 
\renewcommand{\fnum@table}{Table~\thetable}   
\makeatother
\subsection{Collagen network and fibrils properties based on experiments}
\setcounter{table}{0}
\begin{table}[H]
\centering
\caption{Network properties of collagen type I at 4 mg/mL concentration across varying polymerization temperatures (\cite{Jansen2018})} 
\newcolumntype{Y}{>{\centering\arraybackslash}m{0.2\textwidth}}
\newcolumntype{Z}{>{\raggedright\arraybackslash}X}
\begin{tabularx}{\textwidth}{@{}Y >{\centering\arraybackslash}m{0.2\textwidth} Z@{}}
\toprule
\textbf{Property} & \textbf{Temperature} & \textbf{\centering Experimental observations} \\ 
\midrule
\multirow{3}{=}{\textbf{Network appearance}\\ \textit{(Confocal reflectance microscopy)}} 
& 22\textdegree C & Very heterogeneous, open structure with fan-shaped bundles of fibrils. \\
& 26\textdegree C & Heterogeneous, open structure with bundles of fibrils. More uniform bundle width compared to 22\textdegree C. \\
& 30--37\textdegree C & Dense, isotropic, and uniform. \\
\midrule
\multirow{3}{=}{\textbf{Fibrils diameter} \\ \textit{(Scanning electron microscopy  (SEM) and Light scattering (LS))}} 
& 22\textdegree C & \textit{SEM:} Not quantified due to open fan-shaped bundles of fibrils. \\ & & \textit{LS:} Thickest fibrils with an average diameter of 300 nm. \\
& 26--30\textdegree C & \textit{SEM:} 150 nm on average, with significant spread at 26\textdegree C. \\ & & \textit{LS:} Comparable diameters of approximately 200 nm. \\
& 34--37\textdegree C & \textit{SEM:} Consistent diameter around 70 nm. \\ & & \textit{LS:} Smaller diameters of approximately 150 nm. \\
\midrule
\parbox{\linewidth}{\textbf{Fibrils length} \\ \textit{(Light \\scattering)}} 
& 26--37\textdegree C & Reduces from 3.3 \textmu m at 26\textdegree C to 1.6 \textmu m at 37\textdegree C. \\
\bottomrule
\end{tabularx}
\label{tab:CollagenStructure}
\end{table}

\subsection{Comparative analysis of Kullback–Leibler (KL) divergence and the Cramér-von Mises (CVM) test}
To compare the energy computation of CVM with that of KL, we examine runtime and percentage improvement for the two tests across varying connectivity values (z). The runtime for both methods is generally comparable, with CVM marginally outperforming KL in several cases. KL regularly shows larger percentage improvements suggesting more efficiency. The Wasserstein distance (Wass.) is presented as an extra statistic. Overall, KL outperforms CVM in terms of percentage improvement.

\begin{table}[H]
\centering
\caption{Runtime, Improvement Percentage, and Wasserstein Distance: CVM vs KL}
\resizebox{\textwidth}{!}{%
\begin{tabular}{cc|ccc|ccc}
\toprule
\multirow{2}{*}{\textbf{Test}} & \multirow{2}{*}{\textbf{z}} & \multicolumn{3}{c|}{\textbf{CVM Metrics}} & \multicolumn{3}{c}{\textbf{KL Metrics}} \\ \cmidrule(lr){3-5} \cmidrule(lr){6-8}
& & \textbf{Runtime (s)} & \textbf{\% Improvement} & \textbf{Wass.} & \textbf{Runtime (s)} & \textbf{\% Improvement} & \textbf{Wass.} \\ \midrule
1  & 3.12 & 32.07 & 97.53 & 0.0043 & 32.35 & 98.82 & 0.0033 \\
2  & 3.12 & 34.89 & 91.29 & 0.0044 & 35.12 & 99.42 & 0.0033 \\
3  & 3.12 & 30.43 & 87.01 & 0.0044 & 33.80 & 99.65 & 0.0033 \\
4  & 3.12 & 32.38 & 85.64 & 0.0042 & 34.76 & 98.85 & 0.0043 \\
5  & 3.12 & 29.21 & 85.58 & 0.0037 & 32.07 & 99.58 & 0.0026 \\
6  & 3.20 & 34.64 & 95.90 & 0.0066 & 30.35 & 99.15 & 0.0046 \\
7  & 3.20 & 35.81 & 95.69 & 0.0058 & 38.30 & 97.00 & 0.0042 \\
8  & 3.20 & 32.77 & 87.69 & 0.0068 & 37.91 & 99.30 & 0.0055 \\
9  & 3.20 & 31.91 & 96.84 & 0.0064 & 31.85 & 99.32 & 0.0050 \\
10 & 3.20 & 34.43 & 97.44 & 0.0058 & 37.08 & 99.38 & 0.0057 \\
11 & 3.36 & 32.93 & 90.63 & 0.0097 & 30.33 & 97.24 & 0.0067 \\
12 & 3.36 & 34.92 & 89.69 & 0.0062 & 37.27 & 98.74 & 0.0053 \\
13 & 3.36 & 31.02 & 92.83 & 0.0090 & 33.83 & 97.39 & 0.0059 \\
14 & 3.36 & 31.63 & 77.34 & 0.0090 & 30.76 & 94.70 & 0.0077 \\
15 & 3.36 & 33.74 & 86.82 & 0.0067 & 30.36 & 98.69 & 0.0066 \\
16 & 3.56 & 30.13 & 83.92 & 0.0124 & 34.25 & 96.84 & 0.0095 \\
17 & 3.56 & 37.51 & 77.87 & 0.0115 & 30.34 & 96.22 & 0.0099 \\
18 & 3.56 & 34.79 & 85.38 & 0.0098 & 35.87 & 94.90 & 0.0095 \\
19 & 3.56 & 32.04 & 84.11 & 0.0101 & 34.58 & 97.27 & 0.0091 \\
20 & 3.56 & 31.06 & 82.93 & 0.0104 & 31.11 & 95.28 & 0.0103 \\
\bottomrule
\end{tabular}%
\label{tab:CVM_VS_KL}
}
\end{table}

\subsection{Relationship between the number of seeding points and the fiber length}
\begin{figure}[H]
\includegraphics[width=\textwidth]{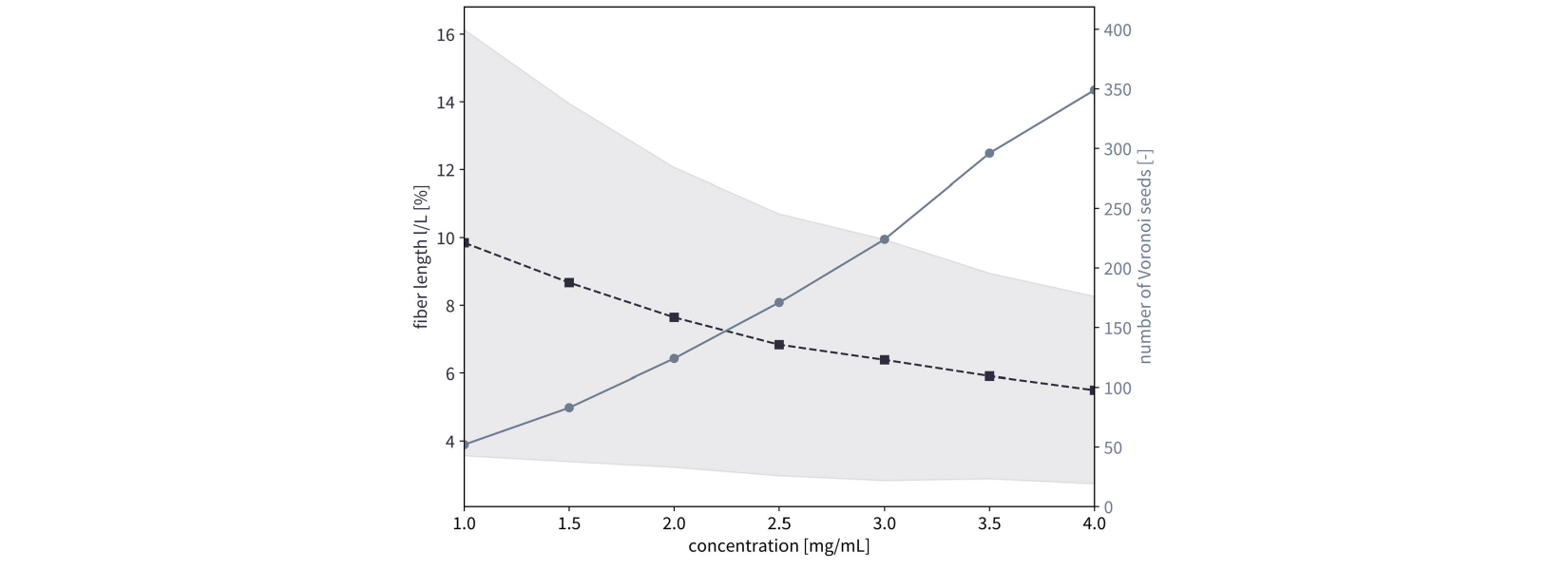}
\caption{\textbf{Fiber characteristic length and number of Voronoi seeds across increasing concentration levels}. 
We generate the 3D Voronoi networks with seed counts refined to match the target density and connectivity of 3.2. 
The left y-axis shows the mean of the normalized fiber length, with shaded areas indicating standard deviation. 
The right y-axis shows the number of Voronoi seeds at each concentration.
}
\label{fig: seeds_length} 
\end{figure}

Figure \ref{fig: seeds_length} shows the average fiber length ($l$) normalized by the domain size ($L$) and corresponding standard deviation after connectivity and length optimization, as a function of concentration. 
We generate seven fiber networks by adjusting the number of Voronoi seeds to match the target concentration while maintaining an average connectivity of 3.2.
Fiber characteristic length ($l/L$) saturates at higher concentrations while Voronoi seeds increase with concentration.
These results highlight that higher seed densities lead to more spatially constrained, finer network structures with shorter connecting fibers. 

\subsection{Network optimization performance}
\begin{figure}[H]
\includegraphics[width=\textwidth]
{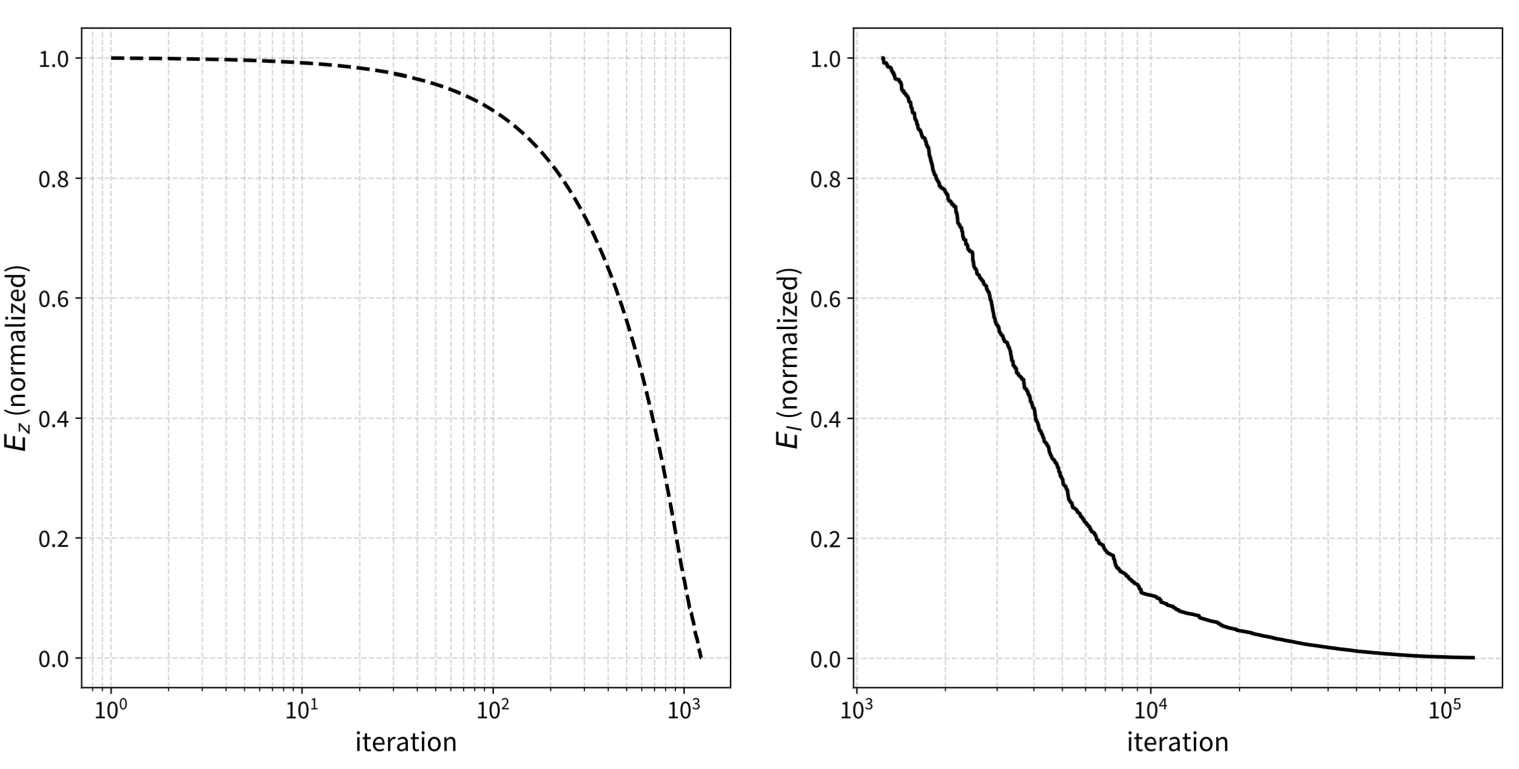}
\caption{\textbf{Convergence check}. 
Performance of the generative pipeline in generating one sample of collagen with a concentration of 4 mg/ml in a cube of edge length 40 \textmu m with average connectivity of 3.12.
The dashed line shows the slow initial trend of the connectivity optimization constrained by the choice of internal edges that must preserve the overall network connectivity while reducing the internal number of cross-links over branches.
The continuum line shows the fast optimization associated with the density-preserving moves to optimize the length distribution in the fiber network.}
\label{fig: performance} 
\end{figure}

We evaluate the performance of our two-step iterative optimization process by analyzing two energy terms: the energy associated with the deviation of the actual connectivity from the target connectivity, denoted $E_z$, and the KL divergence between the actual and target length distributions, denoted $E_l$.
Figure \ref{fig: performance} illustrates the energy output from a representative optimization process.
In the early stages of the dilutive transformation, the algorithm preferentially targets edges inside the RVE domain, removing fibers from internal cross-links to form branches. 
This constrained search initially limits optimization efficiency. 
Performance improves significantly once edge removals accept the boundary edges, which, being periodic, are removed in pairs, leading to enhanced dilution.
After achieving the target connectivity, length optimization begins. At this stage, we quantify the energy associated with each annealing configuration by the KL divergence between the current and target length distributions.

\subsection{Symmetric elastic modulus effect}
\begin{figure}[H]
\includegraphics[width=\textwidth]{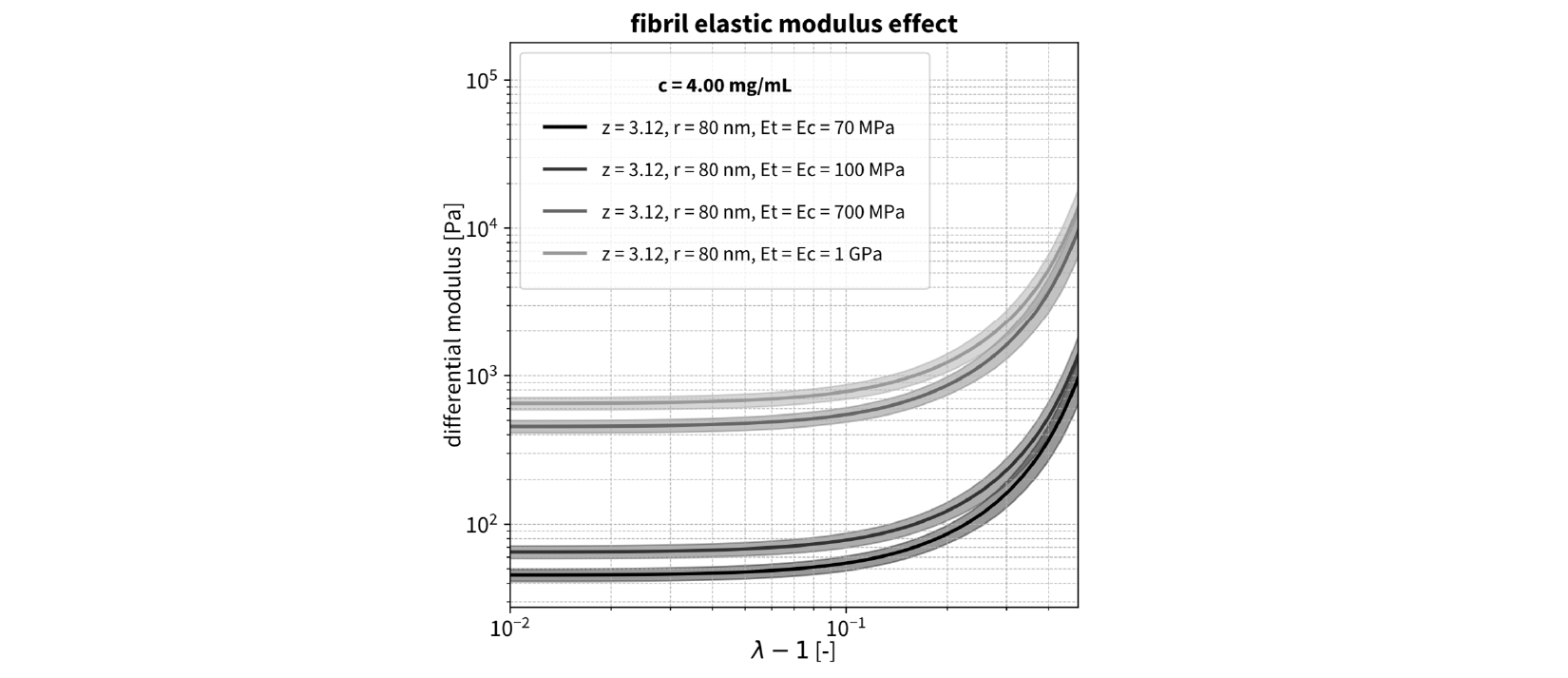}
\caption{\textbf{Effect of fibril stiffening under a single-modulus assumption for both tension and compression.} 
We apply 50\% shear to 10 independent network samples per condition. 
The shaded areas represent the standard deviation across these samples. 
The networks have a concentration of 4 mg/mL, an average connectivity of 3.12, and fibers with an average length of 2 \textmu$\mathrm{m}$ and radius of 0.08 \textmu$\mathrm{m}$. 
The values of the elastic modulus used are consistent with the ranges tested for the bimodulus material model.
}
\label{fig: unimod} 
\end{figure}

Figure \ref{fig: unimod} illustrates the effect of increased fibril elastic modulus under the assumption of tension-compression symmetry.
Here, fibrils are modeled with a simple linear elastic material. 
The differential modulus exhibits a scaling relationship with fibril modulus, which is consistent with theoretical predictions, implying that the macroscopic constitutive response of the network is directly proportional to the constitutive behavior of its constituent fibers.

\end{document}